\documentclass[twocolumn]{aastex7}
\usepackage{amsmath}
\usepackage{enumitem}
\usepackage[normalem]{ulem}

\shorttitle{Free-Floating Planets via Ejection}
\shortauthors{Guo et al.}
\graphicspath{{./}{figures/}}

\begin{document}

\title{Formation of Free-Floating Planets via Ejection: Population Synthesis with a Realistic IMF and Comparison to Microlensing Observations}

\author[orcid=0000-0001-6870-3114,sname='Guo']{Kangrou Guo}
\affiliation{Tsung-Dao Lee Institute, Shanghai Jiao Tong University, 1 Lisuo Road, Shanghai 201210, China}
\email[]{carol.guo@sjtu.edu.cn}  

\author[0000-0001-9564-6186, sname='Ida']{Shigeru Ida}
\affiliation{Earth-Life Science Institute, Institute of Science Tokyo, Meguro, Tokyo 152-8550, Japan}
\affiliation{Department of Astronomy, School of Science, Westlake University, Hangzhou, Zhejiang 310030, China}
\email{ida@elsi.jp}

\author[0000-0002-8300-7990, sname='Ogihara']{Masahiro Ogihara}
\affiliation{Tsung-Dao Lee Institute, Shanghai Jiao Tong University, 1 Lisuo Road, Shanghai 201210, China}
\affiliation{School of Physics and Astronomy, Shanghai Jiao Tong University, 800 Dongchuan Road, Shanghai 200240, China}
\email[]{ogihara@sjtu.edu.cn}

\correspondingauthor{Kangrou Guo, Masahiro Ogihara}
\email{carol.guo@sjtu.edu.cn, ogihara@sjtu.edu.cn}

\begin{abstract}

Microlensing observations suggest that the mass distribution of free-floating planets (FFPs) follows a declining power-law with increasing mass. The origin of such distribution is unclear. Using a population synthesis framework, we investigate the formation channel and properties of FFPs, and compare the predicted mass function with observations. Assuming FFPs originate from planet-planet scattering and ejection in single star systems, we model their mass function using a Monte Carlo based planet population synthesis model combined with N-body simulations. We adopt a realistic stellar initial mass function, which naturally results in a large fraction of planetary systems orbiting low-mass stars.
The predicted FFP mass function is broadly consistent with observation: it follows the observed power-law at higher masses ($10 \lesssim m/M_\oplus < 10^4$), while at lower masses ($0.1 < m/M_\oplus \lesssim 10$) it flattens, remaining marginally consistent with the lower bound of the observational uncertainties.
Low-mass, close-in planets tend to remain bound, while Neptune-like planets at wide orbits dominate the ejected population due to their large Hill radii and shallow gravitational binding.
We also compare the mass distribution of bound planets with microlensing observations and find reasonably good agreement with both surveys.
Our model predicts $\simeq 1.20$ ejected planets per star in the mass range of $0.33 < m/M_\oplus < 6660$, with a total FFP mass of $\simeq 17.98~M_\oplus$ per star. Upcoming surveys will be crucial in testing these predictions and constraining the true nature of FFP populations.

\end{abstract}

\section{Introduction} \label{sec:intro}

Free-floating planets (FFPs) — planetary-mass objects not bound to any host star — represent a unique class of objects that challenge our understanding of planet formation and system dynamics. Recent results from the MOA-II microlensing survey \citep{Sumi_2023} have placed strong constraints on the FFP population, suggesting that their mass distribution follows a power-law with a declining slope toward higher masses. These results build upon earlier work, which indicated the presence of Jupiter-mass FFPs, but the latest survey greatly improves sensitivity at lower masses, unveiling a population dominated by sub-Saturn and Earth-mass objects.

In parallel, microlensing surveys have also provided increasingly robust statistics on bound planets, particularly those at or beyond the snow line. Surveys such as those by \citet{Suzuki_2016} and \citet{Zang_2025} have mapped the distribution of planet-to-star mass ratios across a wide dynamic range. Interestingly, \citet{Zang_2025} reports that the bound planet population exhibits a dip in frequency around a planet-to-star mass ratio of $\log{q} \sim -3.5$ to $-3$, a feature that may encode information about formation and dynamical evolution mechanisms.

The origin of FFPs remains an open question. Several formation channels have been proposed, including direct collapse in isolation akin to brown dwarfs and subsequent photo-erosion \citep[e.g.,][]{Whitworth_2004}, ejection from planetary systems via planet–planet scattering \citep{Veras_2012, Coleman_2025}, perturbations from stellar flybys \citep{Yu_2024}, and dynamical disruption in binary systems \citep{Kaib_2013}. A recent study by \citet{Hadden_Wu_2025_arxiv} has suggested that some microlensing FFPs may, in fact, be bound planets on very wide orbits ($>100$ au), further complicating the interpretation of these detections.

Among these, ejection via planet–planet scattering remains a leading candidate, especially for planets originally formed in protoplanetary disks. However, most studies exploring this scenario rely on idealized N-body setups with ad hoc initial conditions: the number of planets, their spacing, and their mass ratios are often varied arbitrarily as free parameters \citep[e.g.,][]{Veras_2012, Hadden_Wu_2025_arxiv}. This limits the predictive power and interpretability of the resulting FFP statistics.
Circumbinary systems have been proposed as efficient factories of FFPs, and systematic studies of FFP production in such environments have been conducted \citep{Coleman_2024,Coleman_2025}. Yet planetary systems around binaries remain poorly constrained observationally, especially when compared to single-star systems. Moreover, \citet{Coleman_2025} did not account for a stellar mass distribution consistent with the microlensing survey targets when comparing their predicted FFP mass function with the observationally inferred results of \citet{Sumi_2023}.

In this work, we take a different approach. We combine the Ida \& Lin planet population synthesis model (hereafter IL PPS model, \citealp{Ida_2018}) -- which self-consistently tracks disk evolution, planet growth, orbital migration, and orbital instability \citep{Ida_2010, Ida_2013} -- to model the formation and dynamical evolution of planetary systems around single stars. Our initial conditions are not tuned by hand, but instead arise from a physically motivated formation model. To ensure statistical realism, we incorporate a galactic stellar initial mass function (IMF; \citealp{Koshimoto_2021a}), which naturally yields a large fraction of M-dwarf host stars. The shallower gravitational potential of these low-mass stars facilitates ejection of outer planets, making them critical to the FFP population.

In addition, because planet–planet scattering through close encounters is central to the ejection scenario, we assess the robustness of the Monte Carlo prescriptions used in the IL PPS model. By comparing Monte Carlo simulation results with those integrated with full N-body simulations \citep{Guo_2025}, we verify that the statistical outcomes — including the mass function of ejected planets — are consistent between the two approaches. This validation allows us to confidently use the computationally efficient population synthesis results to investigate the statistical properties of both ejected and bound planets under realistic formation and stellar population assumptions.

Our goal is to determine whether planet–planet scattering in single-star systems, modeled with realistic planet formation physics and stellar demographics, can reproduce the mass distribution of free-floating and bound planet populations.
This paper is arranged as follows: in section \ref{sec:method}, we introduce the PPS model and N-body simulations used. In section \ref{sec:results}, we present the simulation results, with a focus on the comparison between MC simulation results and N-body results, as well as the comparison between simulated planet population and microlensing results. Discussions on the results are listed in section \ref{sec:discussion}, and we summarize in section \ref{sec:conslusion}.

\section{Method} \label{sec:method}

\subsection{An overview of the IL model}

We adopt the planet population synthesis framework originally developed in \citet{Ida_2004a}, which models the growth and orbital evolution of planetary embryos within a protoplanetary disk. The formation and evolution of planetary systems are simulated across a range of initial conditions, enabling statistical comparisons with observed exoplanet populations. 
This semi-analytical model accounts for key physical processes including oligarchic growth of planetary cores, gas accretion, orbital migration (Type I and Type II), and dynamical interactions among planets after gas dissipation. 
The disk structure evolves according to a viscously driven, self-similarly decaying profile, and disk parameters are sampled from observationally motivated distributions. 
We adopt a two-$\alpha$ model to account for the low local turbulent viscosity (parameterized as $\alpha_{\rm{turb}}$) in the disk midplane of a wind-driven accretion disk, where magnetorotational instability (MRI) is inactive and forms a ``dead zone,'' while the global accretion efficiency is characterized by a larger parameter $\alpha_{\rm{acc}}$ \citep{Shakura_1973, Ida_2018}.
The type II migration prescription is updated \citep{Guo_2025} to incorporate the global depletion of disk gas due to the accretion of a planet \citep{Tanigawa_2016, Tanaka_2020}, in addition to the gap opening effect \citep{Kanagawa_2018}.

The orbital evolution of planets through scattering of an embryo by another embryo or a giant planet in orbit crossing after the disk phase is calculated in a Monte Carlo approach, which is based on analytical formula and statistical sampling from a certain distribution \citep[e.g.,][]{Ida_2010, Ida_2013}. 
The onset of orbit crossing of planets (orbital instability) due to secular eccentricity increase is approximately estimated.
Ejection of small planets by secular perturbations from gas giants in instability in outer regions 
is also taken into account.
In the presence of disk gas, scattering of small planets by a migrating gas giant 
and resonant trapping between a planet pair in convergent migration are approximately modeled.

This enables the calculation of the long-term evolution of planetary systems for a large number of stars.
For full details, we refer the reader to the original series of papers and their follow-up works \citep[e.g.][]{Ida_2004a, Ida_2004b, Ida_2013, Ida_2018, Guo_2025}.

The key input parameters for our PPS simulations (hereafter referred to as ``MC simulations'') are listed in Appendix \ref{appendix:MC_param}. 
We also discuss the dependence of our results on some important parameters in Section \ref{sec:discussion}.

\subsection{N-body simulations}

To evaluate the robustness of the Monte Carlo prescriptions for orbital instability used in our population synthesis (MC) simulations, we perform direct N-body integrations using initial conditions drawn at the time of disk dispersal, following the procedures of \citet{Guo_2025}. 
For each planet, we adopt the semi-major axis, mass, and eccentricity from the MC output. Inclinations are set to half the eccentricity, and the remaining orbital angles — the longitude of ascending node, argument of pericenter, and mean anomaly — are randomly sampled from a uniform distribution between 0 and $2\pi$.

These parameters serve as the initial conditions for the N-body simulations, which are integrated for $10^8$ years to capture the long-term dynamical evolution of the systems. The final results are then compared against those from the corresponding MC simulations run for the same duration ($t_{\rm{MC,end}}=10^8$ yr).

We use the REBOUND N-body code \citep{Rein_2012}, employing the MERCURIUS integrator. This hybrid scheme combines the symplectic WHFast integrator \citep{Rein_2015} with the high-accuracy, adaptive timestep integrator IAS15 \citep{Rein_2015_IAS15}, allowing for efficient long-term evolution while maintaining accuracy during close encounters. A planet is considered ejected once its heliocentric distance exceeds 10,000 au and its orbital energy becomes positive. 

\subsection{Initial mass function of stars} \label{subsec:IMF}

\begin{figure*}
    \centering
    \includegraphics[width=\textwidth]{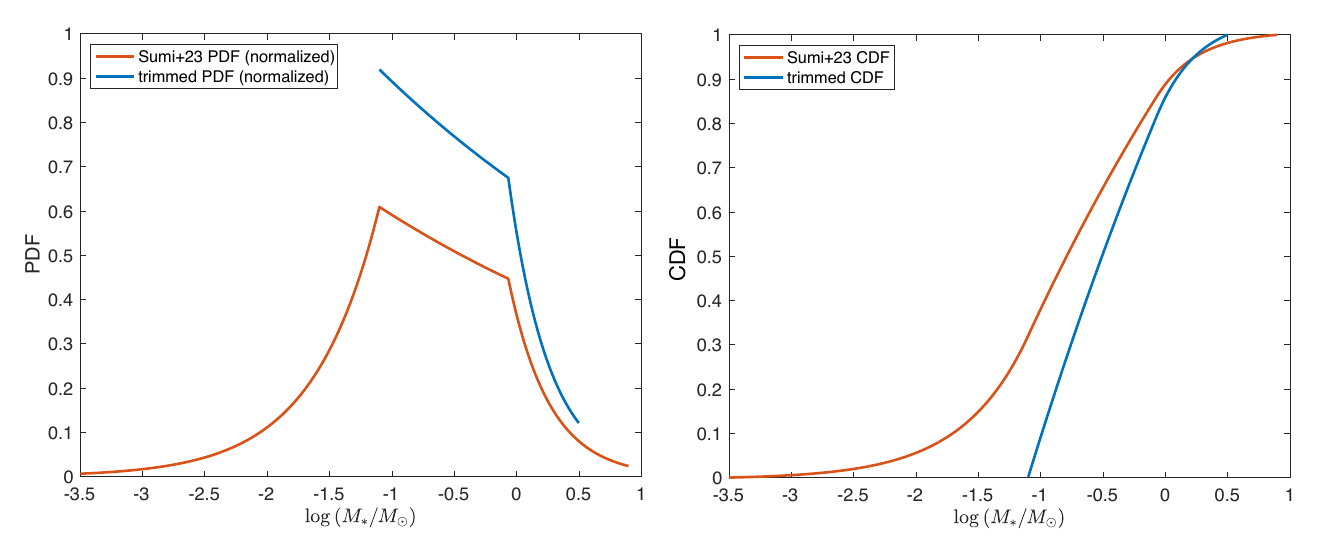}
    \caption{\textit{Left}: normalized probability distribution functions (PDFs) of the stellar IMF used in \citet{Sumi_2023} (red) and in our simulations (blue). \textit{Right}: corresponding cumulative distribution functions (CDFs). Our simulations adopt a trimmed IMF over the range $-1.1 < \log{(M_*/M_\odot)}<0.9$, covering $\simeq 66\%$ of the stellar sample in \citet{Sumi_2023}.}
    \label{fig:stellar_MF}
\end{figure*}

To enable a fair comparison between our simulated and observed mass functions of free-floating planets (FFPs), we adopt a broken power-law initial mass function (IMF) for stars, following \citet{Koshimoto_2021a} and \citet{Sumi_2023}. The IMF is defined as:
\begin{equation}
    \frac{dN}{d\log M_*} \propto 
    \begin{cases}
    M_*^{-\alpha_1} & , M_1 < M_*/M_\odot \leq 120, \\
    M_*^{-\alpha_2} & , 0.08 < M_*/M_\odot \leq M_1, \\
    M_*^{-\alpha_3} & , 3\times 10^{-4} < M_*/M_\odot \leq 0.08.
    \end{cases}
    \label{eq:IMF}
\end{equation}
\citet{Sumi_2023} adopted the E$+\rm{E_X}$ IMF model from \citet{Koshimoto_2021a}, with power-law indices $\alpha_1=1.32$, $\alpha_2=0.13$, $\alpha_3=-0.82$, and a break mass of $M_1=0.86~M_\odot$. They normalized the IMF over the stellar mass range $3\times 10^{-4} < M_*/M_\odot <8$, which includes brown dwarfs and massive stars. 

However, the planet population synthesis (PPS) model we employ is not applicable to very low-mass or high-mass stars. We therefore restrict our sample to stars with masses between 0.08 and $3~M_\odot$, covering approximately 66\% of the stellar mass range used in \citet{Sumi_2023}. 

Figure \ref{fig:stellar_MF} compares the trimmed stellar mass distribution used in our simulations (blue) to that of \citet{Sumi_2023} (red). The left and right panels show the normalized probability distribution functions and cumulative distribution functions respectively. When normalizing the simulated planet mass function for comparison with observationally inferred FFP statistics, we scale our histogram values by a factor of 66\% to account for the reduced mass range.

\subsection{Model setup}

We perform three sets of simulations in this study: 
\begin{enumerate}[label=\Alph*.]
   \item MC simulations for 3000 stars
   \item MC simulations for 300 stars 
   \item N-body simulations (using REBOUND) for 300 stars
\end{enumerate}

For the N-body simulations (set C), we use the planet properties (mass, semi-major axis, and eccentricity) at the time of disk dispersal from simulation set B as initial conditions. Only planets with masses above $0.1~M_\oplus$ and semi-major axes larger than 0.1 au are included. Planets within 0.1 au are excluded because they are deeply embedded in the stellar gravitational potential well and are unlikely to be ejected via dynamical interactions.

Since our focus is on the formation of free-floating planets through planet–planet scattering, and to reduce computational cost, we restrict the N-body sample to planets that are both sufficiently massive and located far enough from the star to be dynamically susceptible to ejection.

\section{Results} \label{sec:results}

\subsection{MC simulation results} \label{subsec:MC_results}

\begin{figure*}
    \centering
    \includegraphics[width=\textwidth]{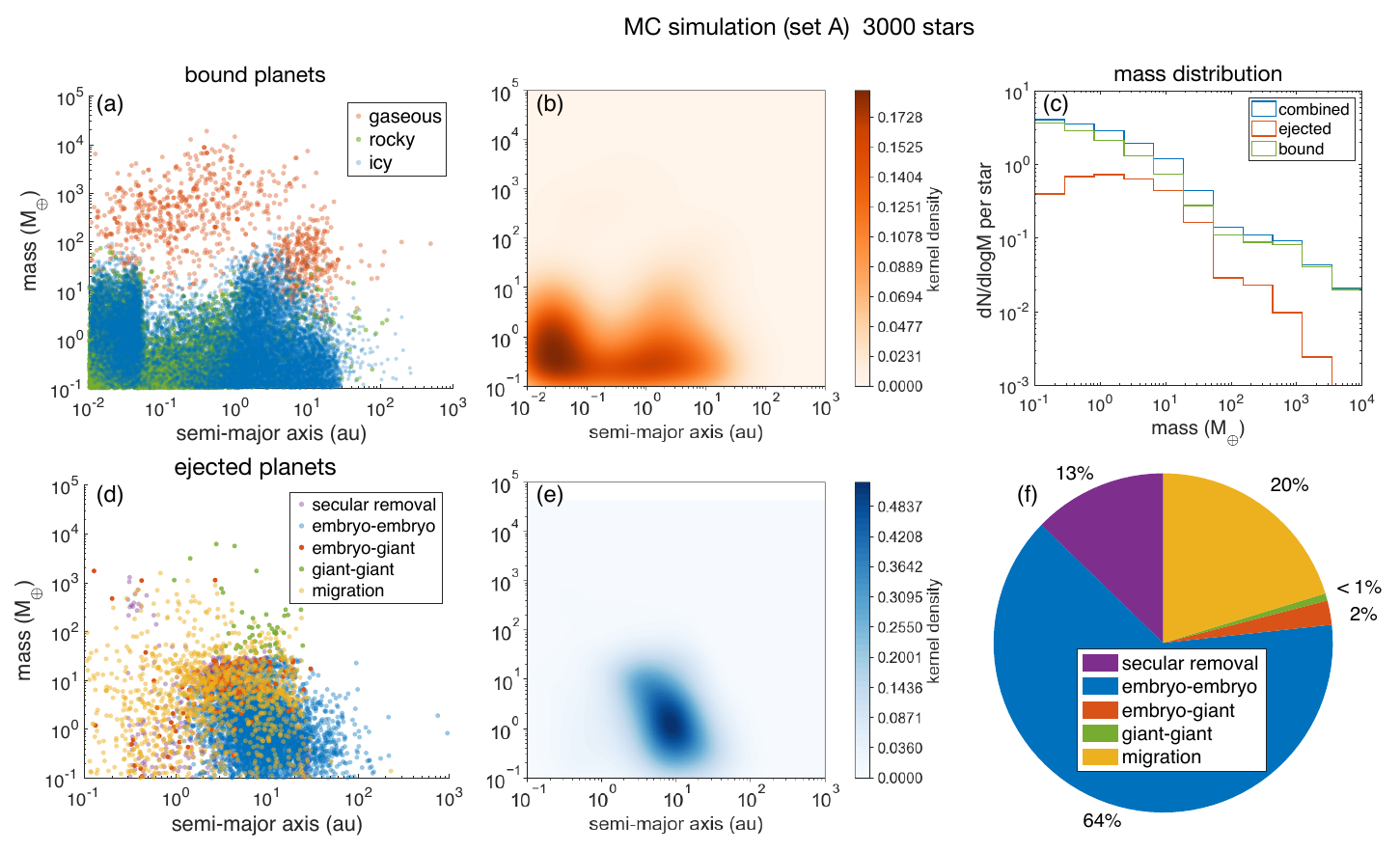}
    \caption{Results of MC simulations for 3000 stars (set A). (a): scatter plot of all bound planets in the 3000 systems at the end of simulation ($t_{\rm{MC,end}} = 10^9$ yr) on the $(a,m)$ plane. Red, blue, and green points represent gaseous, icy, and rocky planets, respectively. (b): 2D KDE of the data points in (a). The color scale represents the KDE values. (c): Mass histograms of the ejected planets (red), bound planets (green), and both population combined (blue). (d): Semi-major axes and masses of ejected planets prior to the moment of ejection. Colors distinguish the causes of ejection of the planets. Purple dots represent small planets (defined as those with $e_{\rm{esc}} < 1$) that are considered removed by enhanced secular perturbations when giant planets become unstable to acquire time-dependent large eccentricities (``secular removal''). Blue dots mark non-giant planets with $m<30~M_\oplus$ (embryos) that are ejected by another embryo during close scattering (``embryo-embryo''). Red dots show embryos that are ejected by giant planets (``embryo-giant''). Green dots indicate giant planets that are ejected by other giant planets (``giant-giant''). Yellow dots show the planets that are ejected during the migration of a giant planet (``secular removal''). (e) Same as (b) but for data points in (c). (f): A pie chart demonstrating the fraction of different ejection channels shown in panel (d).}
    \label{fig:MC_population}
\end{figure*}


Figure \ref{fig:MC_population} (a) shows the bound planet population generated by the MC simulations of 3000 systems (set A) at $t_{\rm{MC,end}} = 10^9$ yr. The model parameters are specified in Appendix \ref{appendix:MC_param}. Gaseous planets, defined as those with total masses exceeding twice their core masses, are marked in red. Icy planets—those with core ice fractions greater than 50\%—are shown in blue, while rocky planets are shown in green. A 2D density plot of the data points in panel (a) is shown in panel (b).

The overall planet distribution reproduces several trends consistent with established planet formation models \citep[e.g.,][]{Ida_2018,Emsenhuber_2021_I}.
For instance, the maximum planet mass increases with orbital distance, reflecting the rise of isolation mass with semi-major axis. Icy planets form beyond the snow line from ice-rich planetesimals, and gas giants emerge just outside the snow line where high solid surface density and moderate accretion timescales allow for rapid core growth and subsequent gas accretion. As a result of efficient type-I migration, many rocky and icy planets drift inward and pile-up within 0.1 au. Gas giants, although typically formed beyond the snow line, also populate the inner region owing to migration. A planet ``desert'' in the  $\simeq 10-100~M_\oplus$ range within a few au is also observed, consistent with earlier studies \citep[e.g.,][]{Ida_2004a}, and attributed to the runaway nature of gas accretion \citep{Pollack_1996}. 

Panel (c) in Figure \ref{fig:MC_population} shows the mass distribution of all the planets (regardless of their distance from the host stars) in the ejected population, the bound population, and two populations combined. When interpreted alongside the $(a, m)$ scatter plots, several trends become evident. Planets with masses between approximately 0.1 and 10 Earth masses tend to remain bound, typically orbiting within $\simeq 10$ au. In contrast, the majority of ejected planets originate around and beyond $\simeq 10$ au and exhibit a relatively flat mass distribution in this small mass range.
At large orbital distances, the feeding zone is wider, allowing planets to grow to higher masses. Many planets beyond $\simeq 10$ au reach Neptune-like masses, with some approaching the threshold for runaway gas accretion but lacking the time to become Jupiter-like giants before the gas disk dissipates. Located in a shallower gravitational potential well, these planets are more vulnerable to dynamical scattering and ejection.
Giant planets with mass above $\simeq 100~M_\oplus$, on the other hand, are less likely to be ejected due to their strong gravitational binding and inertial resistance to perturbations. As a result, the high-mass end of the ejected planet mass distribution is notably suppressed compared to that of the bound or total populations.

Panel (d) of Figure \ref{fig:MC_population} shows the distribution of ejected planets in the $(a,m)$ diagram, where $a$ is the semi-major axis at the moment just before ejection. Panel (e) shows the 2D density plot of the data points in panel (c).
Three ejection mechanisms are identified in the MC simulations:

\begin{enumerate}
    \item Ejection by ``embryo-embryo" (blue), ``embryo-giant" (red), and ``giant-giant" (green) scattering during orbital instability: planet ejection via orbital instability is detected by monitoring the eccentricity (calculated at each timestep with Monte Carlo prescriptions) of each planet after the disk phase. The eccentricity grows from close encounters and scattering between planets. A planet is considered ejected when its eccentricity exceeds unity. 
    
    \item ``Secular removal" (purple) of small planets in inner region by enhanced secular perturbations from gas giants that undergo orbital instability in outer regions:
small planets are considered to be ejected (effectively removed) when giant planet orbital instability happens, owing to violent secular perturbations from highly eccentric giant planets \citep{Matsumura_2013, Ida_2013}.
    ``Small planets'' are those planets without ejection ability, with $e_{\rm{esc}} < 1$, where $e_{\rm{esc}}$ is the ratio of the escape velocity from the planet surface $v_{\rm{esc}}$ and its Keplerian velocity $v_{\rm{K}}$ 
        \begin{equation}
            e_{\rm{esc}} = \frac{v_{\rm{esc}}}{v_{\rm{K}}} = \sqrt{\frac{2Gm/R}{GM_*/a}}.
            \label{eq:e_esc}
        \end{equation}
        We consider small planets with $a>0.3$ au and $e_{\rm{esc}} < 1$ removed, when orbital instability of giant planets with $e_{\rm{esc}} > 1$ occurs. Small planets within 0.3 au are considered difficult to be ejected because they are rather deep in the potential well of the central star.
    \item Ejection by ``migration" (yellow) in disk phase: Planets migrate inward due to tidal interaction with the disk gas. Giant planets can either shepherd or eject inner smaller planets during their migration. When the orbital separation $\Delta a$ between an inner small planet and a giant planet with mass larger than $30~M_\oplus$ becomes smaller than $5h$, where $h$ is the reduced Hill radius $h = [(m_1+m_2)/(3M_*)]^{1/3}$, we evaluate the eccentricity of the inner small planet according to the criteria in \citet{Shiraishi_2008}. If the eccentricity exceeds unity, this planet is considered ejected by the giant planet; if the eccentricity is smaller than unity, it is further damped by the disk gas and the planet moves inward consequently (i.e., shepherding). 
\end{enumerate}

These three channels collectively populate a wide range of ejected planet masses and orbital distances. Panel (f) shows the fraction of contribution of each ejection channel to the total ejection planet population.
Most ejection happens in instability in 
the post-disk phase and is caused by ``embryo-embryo'' scattering (see the blue circles in panel d, the density distribution in panel e, and the blue patch in panel f. Ejection through effective removal of small planets due to giant planet instability (purple) and ejection through migration of giant planets (yellow) also make considerable contribution to the ejected planet population, but the overall fraction is smaller given the low occurrence rate of giant planets around low-mass stars, which is the majority of our stellar sample. 

Although it is often assumed that giant planets are the primary drivers of FFP production since their large masses make them capable of ejecting smaller bodies, our simulations reveal a different picture. We find that Neptune-mass planets, rather than giants, are the dominant contributors to the ejected population. Their prevalence at wide orbits, combined with their relatively low binding energy to the host star, makes them highly susceptible to mutual scattering and eventual ejection. As a result, even low-mass stars, which rarely host gas giants, can still generate substantial numbers of FFPs through the ejection of Neptune-like planets. This explains why adopting a realistic IMF, heavily weighted toward low-mass stars, still yields an abundant FFP population.

\subsection{Comparison with N-body simulation results}

\begin{figure*}
    \centering
    \includegraphics[width=\textwidth]{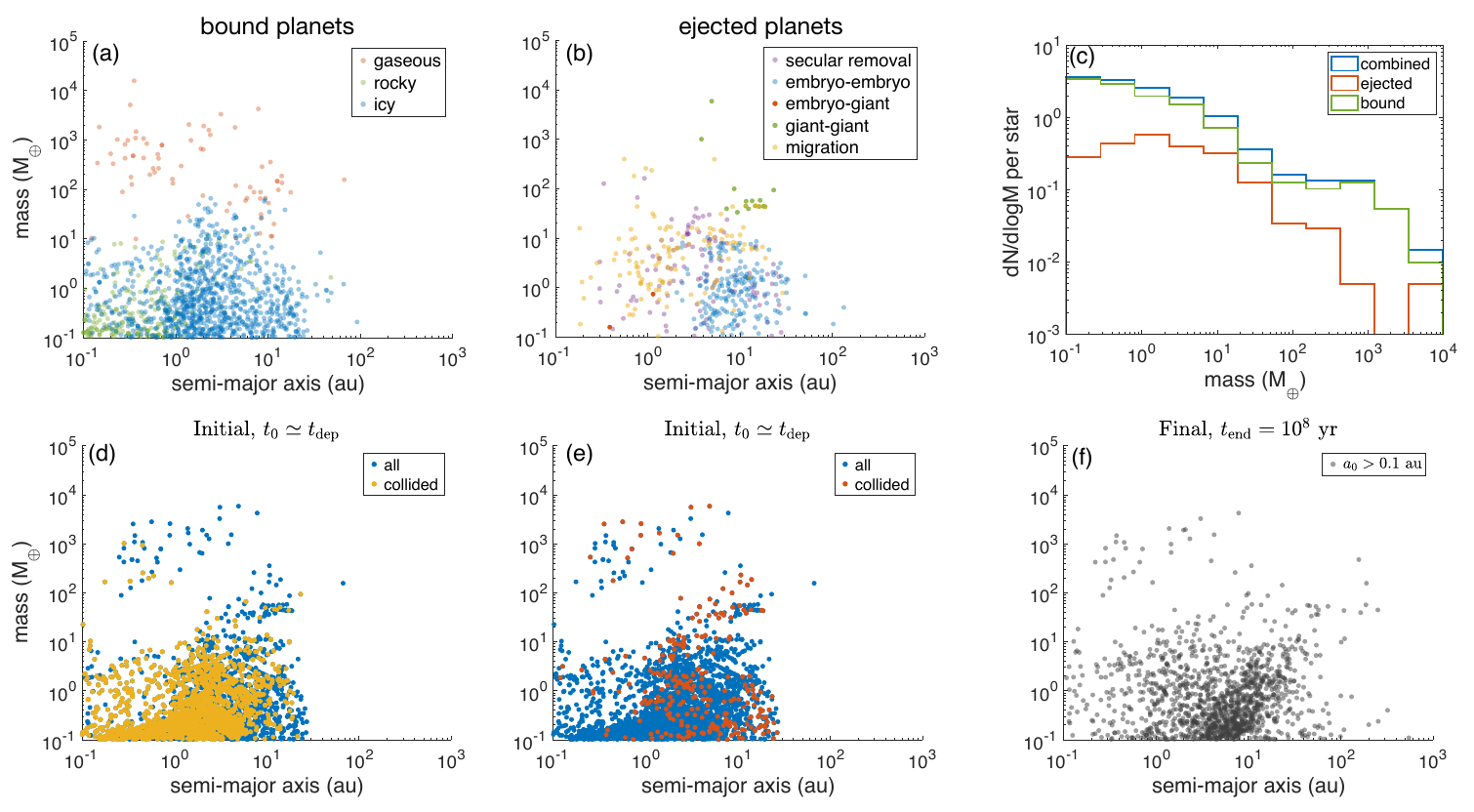}
    \caption{(a)-(c): Same as panels (a), (c), and (e) in Figure \ref{fig:MC_population} but for simulation set B (300 stars). The total simulation time is $t_{\rm{MC,end}}=10^8$ yr. (d)-(e): distribution of planets at $t \simeq t_{\rm{dep}}$, as initial conditions for N-body simulations. The planets that undergo collisions and ejections are highlighted in yellow and red filled circles (d) and (c), respectively. (f): Final planets in REBOUND simulations (at the end of integration $t_{\rm{REB,end}} = 10^8$ yr).}
    \label{fig:MC_300}
\end{figure*}

In this section, we compare the MC simulations with direct N-body simulations to test the robustness of the orbital instability prescriptions used in the MC framework -- an essential component in modeling the production of FFPs.

To ensure a fair comparison, we reduce the sample size of the MC simulations (set B) to 300 stars and limit the integration time to $t_{\rm{MC,end}} = 10^8$ yr. 
Among these 300 systems, 154 form planets with masses greater than $0.1~M_\oplus$, and we perform N-body simulations (set C) for this subset for a total integration time of $t_{\rm{REB,end}} = 10^8$ yr.

Figure \ref{fig:MC_300} shows the $(a,m)$ scatter plots of the bound planets at the end of the MC simulations (a) and the ejected planets at the time of ejection (b). Panel (c) shows the mass distribution of different populations. These results are qualitatively consistent with those in Figure \ref{fig:MC_population}, with reduced data density reflecting the smaller sample size.

Panels (d) and (e) show the planet semi-major axis and mass at the start of N-body simulations ($t \simeq t_{\rm{dep}}$, where $t_{\rm{dep}}$ is the disk depletion timescale). Planets that later collide are marked in yellow, and those that are eventually ejected are shown in red. Panel (f) displays the surviving planets at the end of the simulation (100 Myr), showing their final orbital configurations.

As shown in panel (d), planets that are initially located on close-in orbits are more prone to collisions, as they are located deeper in the gravitational potential well. These small, close-in planets (yellow dot) then merge into larger bodies, populating the region within $\sim$0.1-1 au and $\sim$1-10 $M_\oplus$ on the $(a,m)$ diagram, as seen in panel (f). 
The ejected planets in panel (e) exhibit a similar distribution to those in the MC simulations as in panel (b), though the plotted values correspond to initial conditions rather than pre-ejection orbits.  In both figures we can identify three types of ejected planets:
giant planets ($m \gtrsim 30~M_\oplus$), Neptune-like planets ($m \simeq 1 - 30~M_\oplus$) \footnote{Here by Netpune-like planets, we refer to planets that are mainly represented by the blue dots in Figures \ref{fig:MC_population}(c) and \ref{fig:MC_300}(b), i.e., planets within a few to a few tens of au and within a broad mass range of roughly 1 to $30~M_\oplus$.}, and inner small planets ($m \lesssim 10~M_\oplus$). The inner small planets are mostly ejected during orbital instability of giant planets, as highlighted by the purple dots in panel (b). 
Neptune-like planets located around 10 au dominate the ejected population. Their large Hill radii and relatively weak gravitational binding to the host stars at these wide orbits make them especially susceptible to scattering and ejection. Several giant planets also become ejected after violent scatterings. 

Panel (f) shows the final distribution of bound planets at the end of the N-body simulations. The overall pattern is consistent with that in panel (a) while the data points are in general more dispersed, since N-body simulations capture the mutual interactions of planets more accurately.

\begin{figure*}
    \centering
    \includegraphics[width=\textwidth]{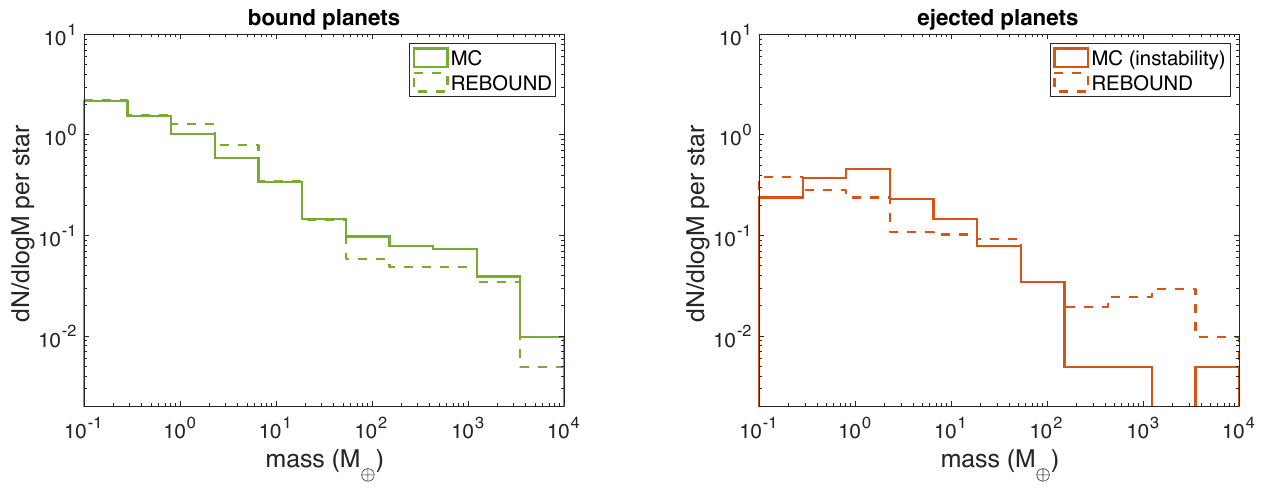}
    \caption{Comparison of the mass distributions of bound planets (left) and ejected planets (right) in MC (solid, set B) and N-body simulations (dashed, set C).}
    \label{fig:compare_mass_function}
\end{figure*}

Next, we compare the mass distributions of the bound and ejected populations resulting from the two simulation methods. Figure \ref{fig:compare_mass_function} shows the mass histograms of the bound planets (left panel) and the ejected planets (right panel) from the MC and N-body simulations respectively.
The MC and N-body results for bound planets agree reasonably well with each other, both showing a decreasing frequency of planets towards the larger masses.
At the higher-mass end, the N-body simulations show a higher fraction of ejected giant planets with masses above $\simeq 100~M_\oplus$ compared to the MC results. 
This discrepancy is likely due to the absence of secular perturbations among giant planets in the IL PPS model.
In the MC simulations (sets A and B), small planets with $e_{\rm{esc}}<1$ are considered ejected when giant planet instability occurs owing to their strong secular perturbations. However, giant planets themselves are only ejected when their eccentricities exceed unity via close encounters. 
In contrast, the N-body simulations capture both short-range scattering and long-range secular interactions among all planets in a system.
Therefore, the N-body simulations produce more ejected giant planets compared with the MC simulations.
Overall, the comparison shows reasonably good agreement between the MC prescriptions and N-body outcomes. For bound planets, the MC and N-body mass histograms match relatively well. For ejected planets, both approaches produce a gently declining mass distribution with increasing planet mass, except at the super giant planet mass end ($m \gtrsim 1000~M_\oplus$). 

Given the consistency between simulation sets B and C, we adopt the more statistically robust results from simulation set A -- which includes a larger sample of 3000 stars and a longer integration time of $t_{\rm{MC,end}} = 10^9$ yr. This longer timescale is particularly important for capturing the evolution of planetary systems around long-lived, low-mass stars.
Further extending $t_{\rm{MC,end}}$ does not significantly alter the results, since most systems have already stabilized and additional instabilities are unlikely.

\subsection{Comparison with observation} \label{subsec:compare_observation}

\begin{figure*}
    \centering
    \includegraphics[width=\textwidth]{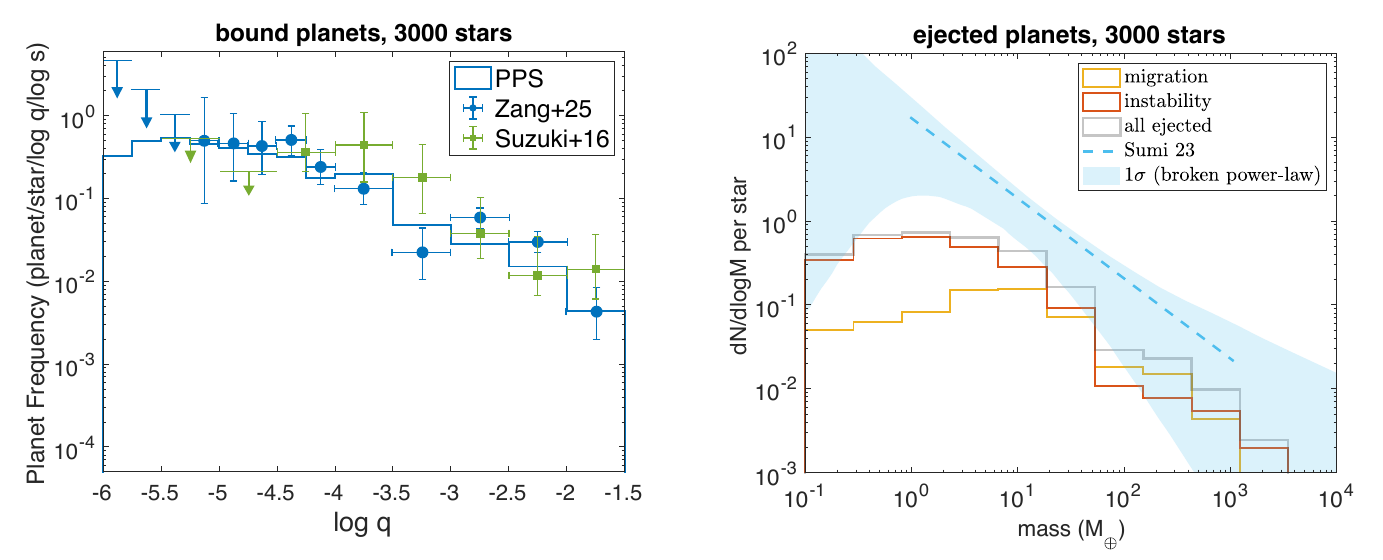}
    \caption{Comparison of MC simulation results with microlensing observations. \textit{Left}: Mass distribution of bound planets (selected from 0.5 to 2 times the Einstein radius $R_E$) from MC simulations. Blue and green data points with error bars show microlensing observation data extracted from \citet{Zang_2025} and \citet{Suzuki_2016}, respectively. \textit{Right}: Mass distribution of ejected planets from MC simulations. The blue dashed line shows the power-law fit from \citet{Sumi_2023}. To cover the larger uncertainties of observational data (especially in the low-mass regime), we plot the $1\sigma$ confidence interval of their broken power-law fit with the blue shaded region. The red and yellow histograms show the planets ejected through orbital instability and migration, respectively. The gray histogram shows the mass distribution of all the ejected planets combining those from orbital instability and migration of giant planets.}
    \label{fig:compare_obs}
\end{figure*}

Using the results from MC simulations A, we compare the mass distributions of both bound and ejected planets against those inferred from microlensing surveys.
The mass distribution of bound planets is compared with observational data from \citet{Suzuki_2016} and \citet{Zang_2025}. Because microlensing is most sensitive to planets located near the Einstein radius of the lens star, we select simulated planets with orbital radii between 0.5 and 2 times the Einstein radius, defined as
\begin{equation}
    R_E = \sqrt{\frac{4GM_*}{c^2} \frac{D_L(D_S-D_L)}{D_S}},
    \label{eq:R_E}
\end{equation}
where $c$ is the speed of light, $G$ is the gravitational constant, $D_L$ is the distance to lens, and $D_S$ is the distance to source. We adopt $D_L = 4$ kpc and $D_S = 8$ kpc, which are representative values for the stars toward the Galactic Bulge.
The left panel in Figure \ref{fig:compare_obs} shows the mass distribution of the bound planets in MC simulations (set A), alongside the microlensing-inferred distributions from \citet{Suzuki_2016} (green) and \citet{Zang_2025} (blue), as a function of the mass ratio $q = m/M_*$. 
Our simulations reproduce the observed distributions within the error bars of both surveys, except in the range  $-3.5 < \log{q} < -3$. The presence of a possible ``dip'' in this regime has been the subject of long-standing debate. Theoretically, the runaway nature of gas accretion can naturally carve such a gap in the mass spectrum, while migration and the dependence of accretion efficiency on orbital radius may obscure or even erase it. In our model, the appearance of the ``dip'', i.e., whether the abundances of giant planets increases for $\log{q}\gtrsim -3$, sensitively depends on the interplay between migration and accretion. We discuss such parameter dependence in more detail in section \ref{subsec:parameter_dependence}.

The right panel of Figure \ref{fig:compare_obs} shows  the comparison between the mass distribution of ejected planets in our MC simulations (set A) and the microlensing-inferred planetary mass function from \citet{Sumi_2023}. 
Overall, the simulation-produced FFP mass distribution match reasonably well with the observationally-inferred values. More specifically, our predicted mass function reproduces the observed declining trend well for planets with masses larger than approximately $10~M_\oplus$. In the lower-mass regime ($m \lesssim 10~M_\oplus$), however, the observational uncertainties are very large, as indicated by the blue shaded region representing the $1\sigma$ error range for the broken power-law fit given in \citet{Sumi_2023}. Within this range, our simulations yield a relatively flat distribution, with histogram values lying near the lower bound of the observational error.
It should be noted that the Earth/super-Earth-mass portion of the mass function is constrained by only about 10 short-timescale events (\citealt{Sumi_2023}), highlighting the limited statistics underlying the current observational fit.
From the theoretical side, as discussed in section \ref{subsec:MC_results}, although the total planet mass function in Figure \ref{fig:MC_population}e (blue, combining bound and ejected populations) shows a monotonic decline with increasing mass, planets with masses between roughly 0.1 and 10 $M_\oplus$ are more likely to remain bound within $\sim$ 10 au. As a result, the mass function of ejected planets (red) in this range appears flatter than that of the overall population. It is primarily the more massive planets around and beyond $\simeq 10$ au that contribute to the ejected population. In this distant region, the planet mass distribution does not show a strong preference toward smaller masses. The reason is two-fold. First, the isolation mass increases with orbital distance, allowing protoplanets at wide separations to continue accreting solid material from a broader feeding zone. On the other hand, since the accretion timescales at large distances are long, many of these planets do not reach the critical mass required for runaway gas accretion soon enough to grow into a Jupiter-like gas giant before the disk dissipates. As a result, there is no pronounced ``planet desert'' in the $\simeq 10$-$100~M_\oplus$ range \citep{Ida_2004a} in the outer disk (see Figure \ref{fig:MC_population}a).
These intermediate-mass planets—residing in a shallower gravitational potential—are more susceptible to ejection and dominate the relatively flat portion of the ejected planet mass function.

\section{Discussion} \label{sec:discussion}

\subsection{The role of giant planets}
\label{subsec:parameter_dependence}


As shown in Figure \ref{fig:MC_population} (panels a and b), giant planets are very rare. This scarcity arises because most stars in our sample are low-mass, and the limited disk mass budget around such stars prevents efficient giant planet formation. As a result, giant planets are not the primary contributors to the FFP population. Figure \ref{fig:MC_population}f shows that, most of the dynamical influence from giant planets manifests in the forms of orbital instabilities that lead to the removal of smaller inner planets (purple) and ejection of planets during migration (yellow). Direct ejections from close encounters with giant planets (green and red) are very rare. More than half of the ejected planets are instead produced by mutual scattering among ``embryos''.

Consequently, the abundance of giant planets only modestly affects the FFP frequency at the low- and high-mass ends of the distribution. For $m\lesssim 10~M_\oplus$, giants enhance the number of FFPs through secular removal of small planets and ejection during migration. For $m\gtrsim 100~M_\oplus$, they can also occasionally eject each other, though this channel is much less efficient. Our simulations suggest that if the occurrence rate of giant planets with $\log{q}\gtrsim -3$ were increased by a factor of $\sim 4$, the frequency of FFPs with $m\lesssim 10~M_\oplus$ would rise by roughly a factor of $\sim 2$.

As mentioned in section \ref{subsec:compare_observation}, the abundance of giant planets in our model is governed by the balance between accretion and migration. The key parameters are the migration efficiency $C_{\rm{mig}}$, and the disk mass budget, which depends on the disk mass scaling factor $f_g$ and the stellar metallicity [Fe/H] (since metallicity sets the solid mass available for core growth). Faster migration reduces the likelihood of giant planet formation, while higher $f_g$ and metallicity increase it. These parameters, however, remain highly uncertain and are only weakly constrained observationally and theoretically. In this work, we adopt ``standard'' values (see Appendix \ref{appendix:MC_param}) that are commonly used in previous population synthesis studies \citep[e.g.,][]{Ida_2018, Guo_2025}.

Despite of the mild influence of giant planet abundance, the overall trend of the FFP mass function predicted by our model remains robust. This is because the dominant population of ejected planets consists of Neptune-like planets with masses $m \lesssim 30~M_\oplus$ located beyond $a\gtrsim 10$ au. These planets reside in dynamically favorable conditions for scattering and are weakly bound to their host stars, making them highly susceptible to ejection. We therefore conclude that, if Neptune-like planets at wide separations are the principal progenitors of FFPs, the mild uncertainties in giant planet frequency in our simulations do not significantly impact the predicted mass function of ejected planets.

\subsection{Comparison with Hadden \& Wu 2025}

\citet{Hadden_Wu_2025_arxiv} propose that a substantial fraction of FFPs are not fully unbound but instead represent a population of ``detached'' planets—objects on wide orbits of a few hundred au. In our classification, planets are considered ejected if their eccentricity exceeds unity ($e>1$) in the MC simulations, or if they simultaneously satisfy the heliocentric distance becoming larger than 10,000 au and orbital energy $E_{\rm{tot}}>0$ in the N-body simulations. To assess the impact of these detached planets to the FFP population, we experimented by including bound planets with $a>100$ au in the FFP mass function. However, this addition produces no significant change in the overall mass function shape, as such planets are relatively rare compared to the population of ``truly'' unbound planets.

It is worth noting that the N-body simulations in \citet{Hadden_Wu_2025_arxiv} span a system age of $10^8$ years, matching the integration time of our REBOUND simulations. Our MC simulations (set A), however, extend to $10^9$ years, providing predictions for more evolved systems. A comparison between the MC results at different system ages reveals that additional Neptune-mass planets are ejected over longer timescales. 
When $t_{\rm{end}}$ is extended from $10^8$ to $10^9$ years, the number of ejected planets roughly doubles within the mass range of $0.1 < m < 10^4~M_\oplus$, and the main increase comes from the planets in the mass range of $0.1 < m \lesssim 30~M_\oplus$.
This gradual contribution shifts the FFP mass function mildly upward in the $m\lesssim 10~M_\oplus$ mass range. 
Therefore, we conclude that our predicted mass function is consistent with the mechanism proposed by \citet{Hadden_Wu_2025_arxiv}.

\subsection{Comparison with Coleman et al. 2025}

Another model that provides predictions for the FFP mass function is presented by \citet{Coleman_2025}, who employ a planet formation framework based on the pebble accretion paradigm. Their results show that FFPs can readily form in circumbinary systems, and several aspects of their predicted mass function are consistent with ours.
First, they find a dearth of ejected planets in the Earth-mass regime, with generally low ejection frequencies for planets below $\simeq 10~M_\oplus$ (see their Figure 4). In addition, they find that for FFPs with masses beyond $10~M_\oplus$, their mass function aligns well with the power-law fit inferred from microlensing by \citet{Sumi_2023}. Both of these trends agree with our simulation results.
However, their predicted mass function for single-star systems yields very low planet frequencies -- below 0.1 planets per star per 0.2 dex in mass -- despite reproducing a broadly monotonic decline toward higher masses. In the case of binary-star systems, they identify a trough in the terrestrial-mass range and a peak near $8~M_\oplus$. This structure arises from planets that reach pebble isolation mass, migrate toward the central cavity, and are subsequently ejected by interactions with the central binary. 
Migration of resonant chains plays an important role in this process: as multiple planets become locked in mean motion resonances and migrate inward together, they can push one another into the circumbinary cavity and trigger ejection, preferentially removing low-mass planets \citep{Fitzmaurice_2022}. Our simulations incorporate resonant chain migration through a simplified analytical prescription \citep{Ida_2010, Ida_2013}, which neglects ejection during orbit crossing prior to merging in the case of resonant chains of Earth/super-Earth-mass planets \footnote{At wider orbits around low-mass stars, this approximation may underestimate the ejection of Earths/super-Earths during resonant chain migration.}. In single-star systems, however, crossings near the inner disk edge do not lead to ejection, since the edge lies very close to the host star and there is no unstable cavity region as in the circumbinary case.


In contrast, our simulations predict higher frequencies of low-mass ejected planets ($m < 10~M_\oplus$) compared to both the single and binary models of \citet{Coleman_2025}. One important distinction is the integration timescale: their simulations are limited to 10 Myr, while ours extend to 1 Gyr or more, enabled by the computational efficiency of the Monte Carlo method. This longer timespan captures additional dynamical evolution, allowing more planets -- particularly Neptune-mass objects at wide separations -- to be gradually scattered and ejected without requiring binary-induced perturbations.
Moreover, their simulations do not explore a range of stellar masses, whereas our model incorporates a realistic stellar IMF based on \citet{Koshimoto_2021a} and \citet{Sumi_2023}. Stellar mass plays a crucial role in the production of FFPs via planet–planet ejection, as it directly influences the gravitational binding energy of planets. By incorporating a realistic stellar mass distribution, our simulations offer a more representative sampling of stellar environments and enable a fairer comparison with microlensing survey results.

It is important to note that the planet population in circumbinary systems remains largely unconstrained observationally \citep[see review by][]{Kostov_2023}, making it difficult to calibrate formation and ejection models in that context. 
The CBPs detected to date are all Neptune-sized or larger, and it remains unclear whether smaller planets exist in such systems; their apparent absence could itself point to efficient ejection of lower-mass planets.
In contrast, single-star planetary systems have been extensively characterized by both microlensing and other detection methods, providing a critical empirical baseline. A key strength of our approach is that the same self-consistent single-star model simultaneously reproduces both the bound and ejected planet populations, providing greater confidence in its predictions for the FFP mass function.

\subsection{Predictions for FFP observations}

Assuming that FFPs form primarily via planet–planet ejection in single-star systems, and within the parameter ranges explored in our simulations, our results suggest that the ejected-planet mass function reproduces the observed power-law decline at $m \gtrsim 10~M_\oplus$, as inferred from the MOA-II survey \citep{Sumi_2023}. At lower masses $(0.1 < m/M_\oplus \lesssim 10)$, our simulations yield a relatively flat distribution, with abundances lying near the lower bound of the $1\sigma$ range of their broken power-law fit.
Based on this mass function, we estimate that there are, on average, approximately 1.20 ejected planets per star within the mass range of $0.33 < m/M_\oplus < 6660$, contributing a total mass of $\simeq 17.98~M_\oplus$ per star. These values are substantially lower than those inferred from the power-law mass function in \citet{Sumi_2023}, which suggest around $21^{+23}_{-13}$ FFPs per star and a total mass of $\simeq 80^{+73}_{-47}~M_\oplus$ over the same mass interval. 
The discrepancy is dominated by the uncertain low-mass regime, where microlensing constraints are weakest. Restricting the comparison to $10 \lesssim m/M_\oplus < 6660$, our results yield $\sim 0.22$ FFP per star with a total mass of $\sim 15.53~M_\oplus$, compared to $\sim 0.79$ FFP per star and $\sim 56.74~M_\oplus$ from the power-law fit. Within this better-constrained mass range, the differences between theory and observation are significantly reduced.

In summary, our results indicate broad consistency with microlensing constraints at Neptune to Saturn masses, with residual tension confined to the least certain low-mass end.
Because the present microlensing constraints on Earth/super-Earth-mass objects are based on just about 10 events \citep{Sumi_2023}, the uncertainties remain substantial. 
We anticipate that future observations, e.g., the upcoming Roman Space Telescope and Earth 2.0, would increase the sample size by orders of magnitude, enabling a transformation of our knowledge of the FFP population comparable to the leap from ground-based transits to Kepler.
These improved statistics will place much tighter constraints on the FFP mass function, and will be essential for distinguishing between different formation pathways and testing the validity of ejection-based models such as the one presented here.

\section{Conclusions} \label{sec:conslusion}

We investigated the formation of FFPs by studying the mass function of FFPs produced via planet–planet scattering in single-star systems, using a combination of population synthesis and N-body simulations. Our models incorporate a realistic stellar initial mass function and are grounded in a self-consistent planet formation framework, avoiding ad hoc initial conditions commonly adopted in earlier studies on planet-planet scattering.

We find that the mass distribution of ejected planets is broadly consistent with recent microlensing observations, especially in the mass range of $10 \lesssim m/M_\oplus < 10^4$. For smaller planets below $10~M_\oplus$, our simulation estimates depart from the steep single power-law slope and fall marginally within the $1~\sigma$ uncertainty range of the broken power-law mass function in \citet{Sumi_2023}.
Over the mass range $0.33 < m/M_\oplus < 6660$, we predict an average of $\sim 1.20$ ejected planets per star, with a total FFP mass of $\sim 17.98~M_\oplus$ per star. Restricting the comparison to the better-constrained $10 \lesssim m/M_\oplus < 6660$ range, the discrepancy is substantially reduced. Overall, our results indicate broad consistency with microlensing constraints, with remaining tension confined to the uncertain low-mass end.

Most ejected planets in our simulations are Neptune-like objects located beyond 10 au, where the combination of shallow gravitational binding and wide feeding zones promotes both growth and ejection. In contrast, lower-mass planets tend to remain bound, while gas giant planets are too massive to be ejected efficiently. Comparison with N-body simulations confirms that the orbital instability prescriptions used in the population synthesis model produce statistically robust results. 

Our results suggest that the true FFP population may be dominated by intermediate-mass planets formed at wide separations, rather than large numbers of terrestrial-mass bodies. The mass function predicted here provides a testable benchmark for upcoming surveys -- particularly with the Roman Space Telescope -- which will be capable of probing both the frequency and mass spectrum of bound and unbound planetary-mass objects with unprecedented sensitivity.

\begin{acknowledgments}
We thank Prof. Yanqin Wu for valuable discussions and comments. We thank the anonymous reviewer for their constructive comments, which helped improve the quality of this manuscript.
This work is supported by National Natural Science Foundation of China (grant No. 12503069).
K. G. acknowledges the support from K. C. Wong Educational Foundation.
S. I. is supported by JSPS Kakenhi grant 21H04512 and 23H00143.
M. O. is supported by National Natural Science Foundation of China (grant No. 12273023).
Numerical computations were carried out on the general-purpose PC cluster at the Center for Computational Astrophysics, National Astronomical Observatory of Japan.
\end{acknowledgments}


\software{Matlab, Matplotlib, WebPlotDigitizer}

\appendix

\section{MC simulation parameters} \label{appendix:MC_param}

In Table \ref{tab:MC_param}, we list the key input parameters used in MC simulation sets A and B. The stellar masses are drawn from the IMF described in Section \ref{subsec:IMF}. The metallicity ([Fe/H]), the logarithm of the disk depletion timescale ($t_{\rm{dep}}$), and the logarithm of the disk mass scaling factor ($f_g$) are sampled from normal distributions with the mean and standard deviation values specified in the first and second columns of the table, respectively.
The gas accretion timescale is parameterized as the Kelvin-Helmholtz timescale $\tau_{\rm{KH}} = 10^x~(m/M_\oplus)^{-y}$ yr, with $x=9$ and $y=3.5$. 
The disk viscosity is described using two parameters: $\alpha_{\rm{acc}}$, which governs global accretion, and $\alpha_{\rm{turb}}$, which represents local turbulence. Following \citet{Ida_2018}, we adopt $\alpha_{\rm{turb}} = 0.1 \alpha_{\rm{acc}}$ and set $\alpha_{\rm{acc}} = 3\times 10^{-3}$.
The inner and outer edges of the planetesimal disk are set to $a_{\rm{min}} = 0.05$ au and $a_{\rm{out}} = 30$ au. The initial gas surface density follows the radial profile of a steady accretion disk with constant $\alpha$ viscosity: $\Sigma_g = \Sigma_{g1}f_g (r/1~{\rm{au}})^{q_g}$, where $\Sigma_{g1} = 750~\rm{g}~cm^{-2}$ is the surface density at 1 au, and $f_g$ is the disk mass scaling factor.
The solid surface density is based on the minimum mass solar nebula (MMSN; \citealp{Hayashi_1981}): $\Sigma_d = \Sigma_{d1}f_d \eta_{\rm{ice}}(r/1~{\rm{au}})^{q_d}$
with $\Sigma_{d1} = 10~\rm{g}~cm^{-2}$, $f_d$ as the solid mass scaling factor, and $\eta_{\rm{ice}}$ as the ice enhancement factor. We set $\eta_{\rm{ice}} = 0$ for $r < a_{\rm{ice}}$ and $\eta_{\rm{ice}} = 4.2$ for $r \geq a_{\rm{ice}}$, where the snowline is defined as $a_{\rm{ice}} = 2.7(M_*/M_\odot)$.
In the standard MMSN model, $f_g = f_d = 1$. For stars of varying metallicity, we assume $f_d = f_g\cdot [\rm{Fe/H}]$, and the overall disk mass scales with stellar mass as $f_g \propto (M_/M_\odot)^{p_{\rm{disk}}}$.
To account for possible reduction of Type I migration rate due to non-linear effects in the disk, we adopt a migration efficiency factor \citep{Ida_2008a} of $C_{\rm{mig}} = 0.1$, effectively increasing the migration timescale (given by \citealp{Tanaka_2002}) by a factor of 10. The total simulation timespan is $t_{\rm{end}}$ is $10^9$ years.
All other parameters follow standard choices from prior works using the IL PPS model.

\begin{deluxetable*}{ccc}
\tablewidth{0pt}
\tablecaption{Parameters of MC simulations} \label{tab:MC_param}
\tablehead{
\colhead{Parameter} & \multicolumn{2}{c}{\centering values}
}
\startdata
$N_{\rm{star}}$ & 3000 (A) & 300 (B) \\
$[\rm{Fe}/H]$, $\mathcal{N}(\mu, \sigma^2)$ & $\mu=0.0$ & $\sigma=0.3$ \\
$\log{t_{\rm{dep}}}$, $\mathcal{N}(\mu, \sigma^2)$ & $\mu=6.5$ & $\sigma=0.5$ \\
$\log{f_g}$, $\mathcal{N}(\mu, \sigma^2)$ & $\mu=0.0$ & $\sigma=1.0$ \\
$\tau_{\rm{KH}} = 10^x~(m/M_\oplus)^{-y}$ & $x=9.0$ & $y=3.5$ \\
$\alpha_{\rm{acc}}$, $\alpha_{\rm{turb}}$ & $3\times 10^{-3}$ & $3\times 10^{-4}$ \\
$a_{\rm{min}}$, $a_{\rm{max}}$ (au) & 0.05 & 30 \\
$q_d$, $q_g$ & -1.5 & -1.0 \\
$p_{\rm{disk}}$ & \multicolumn{2}{c}{\centering 1.0} \\
$C_{\rm{mig}}$ & \multicolumn{2}{c}{\centering 0.1} \\
$\log{t_{\rm{end}}}$ & 9.0 & 8.0 \\
\enddata
\tablecomments{The metallicity $[\rm{Fe}/H]$, the logarithm of the disk lifetime $\log{t_{\rm{dep}}}$, and the logarith of the disk mass scaling factor $\log{f_g}$ are each drawn from normal distributions characterized by a mean $\mu$ and dispersion $\sigma$. The disk viscosity is characterized by two parameters: $\alpha_{\rm{acc}}$ for global accretion and $\alpha_{\rm{turb}}$ for local turbulence. Planetesimals are initially distributed between $a_{\rm{min}}$ and $a_{\rm{max}}$. The surface densities of solids and gas follow power-law profiles with indices $q_d$ and $q_g$, respectively. The disk mass scales with stellar mass as $f_g \propto (M_*/M_\odot)^{p_{\rm{disk}}}$. The parameter $C_{\rm{mig}}$ is the retardation factor applied to Type I migration efficiency. Simulations set A spans $10^9$ years, while set B is limited to $10^8$ to allow direct comparison with N-body results.}
\end{deluxetable*}


\bibliography{references}{}
\bibliographystyle{aasjournalv7}


\end{document}